\documentclass[letter,longauth]{aa} 
\usepackage{graphicx}
\usepackage{txfonts}
\def\deg{\ifmmode{^\circ} \else {$^\circ$} \fi}
\def\arcmin{\ifmmode{^\prime} \else {$^{\prime}$} \fi}
\def\arcsec{\ifmmode{^{\prime \prime}} \else {$^{\prime \prime}$} \fi}

\begin{document}
   \title{
{\it Herschel}-PACS spectroscopy of the intermediate mass protostar NGC7129 FIRS 2
\thanks{{\it Herschel} is an ESA space observatory with science instruments provided by European-led Principal Investigator consortia and with important participation from NASA
}
}
\titlerunning{PACS spectroscopy of NGC 7129 FIRS2}

\author{M. Fich\inst{1}
\and D. Johnstone\inst{2,3}
\and T.A. van Kempen\inst{4,5}
\and C. M$^{\textrm c}$Coey\inst{1,6}
\and A. Fuente\inst{7}
\and P. Caselli\inst{8,31}
\and L.E. Kristensen\inst{3}
\and R. Plume\inst{9}
\and J. Cernicharo\inst{10}
\and G.J. Herczeg\inst{11}
\and E.F. van Dishoeck\inst{4,11}
\and S. Wampfler\inst{12}
\and P. Gaufre\inst{13}
\and J.J. Gill\inst{14}
\and H. Javadi\inst{14}
\and M. Justen\inst{15}
\and W. Laauwen\inst{16}
\and W. Luinge\inst{16}
\and V. Ossenkopf\inst{15}
\and J. Pearson\inst{14}
\and R.~Bachiller\inst{7}
\and A.~Baudry\inst{13}
\and M.~Benedettini\inst{19,20}
\and E.~Bergin\inst{20}
\and A.O.~Benz\inst{12}
\and P.~Bjerkeli\inst{17}
\and G.~Blake\inst{21}
\and S.~Bontemps\inst{13}
\and J.~Braine\inst{13}
\and S.~Bruderer\inst{12}
\and C.~Codella\inst{19}
\and F.~Daniel\inst{22,23}
\and A.M.~di~Giorgio\inst{19}
\and C.~Dominik\inst{24}
\and S.D.~Doty\inst{25}
\and P.~Encrenaz\inst{26}
\and T.~Giannini\inst{19}
\and J.R.~Goicoechea\inst{10}
\and Th.~de~Graauw\inst{16}
\and F.~Helmich\inst{16}
\and F.~Herpin\inst{13}
\and M.R.~Hogerheijde\inst{4}
\and T.~Jacq\inst{13}
\and J.K.~J{\o}rgensen\inst{27}
\and B.~Larsson\inst{28}
\and D.~Lis\inst{21}
\and R.~Liseau\inst{17}
\and M.~Marseille\inst{16}
\and G.~Melnick\inst{4}
\and B.~Nisini\inst{19}
\and M.~Olberg\inst{17}
\and B.~Parise\inst{29,30}
\and C.~Risacher\inst{16}
\and J.~Santiago\inst{7}
\and P.~Saraceno\inst{19}
\and R.~Shipman\inst{16}
\and M.~Tafalla\inst{7}
\and F.~van der Tak\inst{16,18}
\and R.~Visser\inst{4}
\and F.~Wyrowski\inst{29}
\and U.A.~Y{\i}ld{\i}z\inst{4}
}

\institute{
Department of Physics and Astronomy, University of Waterloo,
Waterloo, Ontario, Canada N2L 3G;
\email{fich@uwaterloo.ca}
\and
National Research Council Canada, Herzberg
Institute of Astrophysics, 5071 West Saanich Rd, Victoria, BC, V9E
2E7, Canada
\and
Department of Physics \& Astronomy, University of Victoria,
Victoria, BC, V8P 1A1, Canada
\and
Leiden Observatory, Leiden University, P.O. Box 9513, 2300 RA Leiden, 
The Netherlands
\and
Harvard-Smithsonian Center for Astrophysics, 60 Garden Street, MS 42, Cambridge, MA 02138, USA
\and
Department of Physics and Astronomy, the University of Western Ontario, 
London, Ontario, Canada, N6A 3K7
\and
IGN Observatorio Astron\'{o}mico Nacional, Apartado 1143, 28800 Alcal\'{a} de Henares, Spain
\and
School of Physics and Astronomy, University of Leeds, Leeds LS2 9JT, UK 
\and
Department of Physics and Astronomy, University of Calgary,
Calgary, Alberta, Canada 
\and
Department of Astrophysics, CAB, INTA-CSIC, Crta Torrej\'{o}n a Ajalvir km 4, 28850 Torrej\'{o}n de Ardoz, Spain
\and
Max-Planck-Institut f\"ur extraterrestrische Physik, Garching, Germany
\and
Institute of Astronomy, ETH Z\"urich, 8093 Z\"urich, Switzerland
\and
Universit\'{e} de Bordeaux, Laboratoire d¿Astrophysique de Bordeaux, France; CNRS/INSU, UMR 5804, Floirac, France
\and
Jet Propulsion Laboratory, California Institute of Technology, Pasadena, CA 91109, USA
\and
KOSMA, I. Physik. Institut, Universität zu K\"oln, Zülpicher Str. 77, D 50937 K\"oln
\and
SRON Netherlands Institute for Space Research, Landleven 12, 9747 AD Groningen
\and
Department of Radio and Space Science, Chalmers University of Technology, Onsala Space Observatory, 439 92 Onsala, Sweden
\and
Kapteyn Astronomical Institute, University of Groningen, PO Box 800, 9700 AV, Groningen, The Netherlands
\and
INAF - Istituto di Fisica dello Spazio Interplanetario, Area di Ricerca di Tor Vergata, via Fosso del Cavaliere 100, 00133 Roma, Italy
\and
Department of Astronomy, The University of Michigan, 500 Church Street, Ann Arbor, MI 48109-1042, USA
\and
California Institute of Technology, Division of Geological and Planetary Sciences, MS 150-21, Pasadena, CA 91125, USA
\and
Observatoire de Paris-Meudon, LERMA UMR CNRS 8112, 5 place Jules Janssen, 92195 Meudon Cedex, France
\and
Department of Molecular and Infrared Astrophysics, Consejo Superior de Investigaciones Cientificas, C/ Serrano 121, 28006 Madrid, Spain
\and
Astronomical Institute Anton Pannekoek, University of Amsterdam, Kruislaan 403, 1098 SJ Amsterdam, The Netherlands 
\and
Department of Physics and Astronomy, Denison University, Granville, OH, 43023, USA
\and
LERMA and UMR 8112 du CNRS, Observatoire de Paris, 61 Av. de l'Observatoire, 75014 Paris, France
\and
Centre for Star and Planet Formation, Natural History Museum of Denmark, University of Copenhagen,{\O}ster Voldgade 5-7, DK-1350 Copenhagen, Denmark
\and
Department of Astronomy, Stockholm University, AlbaNova, 106 91 Stockholm, Sweden
\and
Max-Planck-Institut f\"{u}r Radioastronomie, Auf dem H\"{u}gel 69, 53121 Bonn, Germany
\and
Physikalisches Institut, Universit\"{a}t zu K\"{o}ln, Z\"{u}lpicher Str. 77, 50937 K\"{o}ln, Germany
\and
INAF - Osservatorio Astrofisico di Arcetri, Largo E. Fermi 5, 50125 Firenze, Italy
 }

   \date{\bf Accepted May 15, 2010 }

 
  \abstract
   {}
   {
We present preliminary results of the first Herschel spectroscopic observations of NGC7129 FIRS2, an intermediate mass star-forming region.  
We  attempt to interpret the observations in the framework of an in-falling spherical envelope.
    }
   {
The PACS instrument was used in line spectroscopy mode (R=1000-5000) with 15 spectral bands between 63 and 185 $\mu$m. 
This provided good detections of 26 spectral lines seen in emission, including lines of H$_2$O, CO, OH, O~I, and C~II. 
  }
   {
Most of the detected lines, particularly those of H$_2$O and CO, are substantially stronger than predicted by the spherical envelope models, typically by several orders of magnitude. 
In this paper we focus on what can be learned from the detected CO emission lines.
   }
   { 
It is unlikely that the much stronger than expected line emission arises in the (spherical) envelope of the YSO.  
The region hot enough to produce such high excitation lines within such an envelope is too small to produce the amount of emission observed. 
Virtually all of this high excitation emission must arise in structures such as as along the walls of the outflow cavity with the emission produced by a combination of UV photon heating and/or non-dissociative shocks. 
}


   \maketitle
%

\section{Introduction}

The common paradigm for the formation of low mass stars includes the spherical infall of a cloud core, the formation of a disk with a protostar at its centre, and a bipolar outflow emitted from the inner region of this disk/protostar structure. 
Large (R $\sim$ 10,000 AU), low temperature (T $\sim$ 20 K) envelopes, which are often approximated to be nearly spherical (Terebey, Shu, and Cassen \cite{terebey84}) are observed around young low mass stars  and dominate the observed sub-millimeter emission (Shirley et al \cite{shirley00}). 
Recent observations show additional structures, possibly hot outflow cavity walls heated by UV photons (van Kempen et al \cite{vankempen09}) and shocks (Gianninni et al \cite{giannini99}, van den Ancker et al \cite{vandenancker00}, Nisini et al \cite{nisini02}, Arce et al \cite{arce07}, van Kempen et al \cite{vankempen10}). 
Our understanding of the formation of high mass stars is not so mature, although it is clear that there are some significant differences between the low mass and high mass formation mechanisms, including the fact that massive stars are predominantly born in clusters {where interactions between stars are important in the formation mechanism. Low mass stars likely form in clusters too, but their early evolution may be unaffected by neaerby stars until when the first high mass star forms in their vicinity. Then the formation process for the low mass stars probably changes dramatically and perhaps even ceases}. 
The many unanswered questions about these mechanisms include: is there a continuum of properties between the low and high mass young stellar object; where is the boundary between low mass and high mass stars; do these ``intermediate mass (IM)'' stars form singly or in clusters? 

The scarcity of IM stars has made answering such questions difficult. 
High mass stars are also rare in the vicinity of the Sun but their high luminosities make it possible to detect and study them at much larger distances.  
Intermediate mass protostars are usually defined as those Young Stellar Objects (YSOs) with bolometric luminosities between 75 L$_\odot$ and $2\times 10^3$ L$_\odot$, although both the lower and upper limits of this range vary slightly from one author to another.  
There are no examples of well-studied very young protostars (e.g. Stage 0/Class 0) with luminosities within this range within 400 parsecs of the Sun. 
There are a handful of such objects with distances between 400 and 1500 pc, which are individually identified and not confused in larger OB associations.

The recent launch of the ESA {\it Herschel Space Observatory} (Pilbratt et al \cite{pilbratt}) provides us with an opportunity to study IM stars with previously unavailable molecular probes.  
The Water In Star forming regions with {\it Herschel} (WISH) Key Program uses emission from water and other key molecules to probe the conditions in a variety of star forming regions.  

At a distance of $1260\pm50$ parsecs (Shevchenko \& Yakubov, \cite{shevchenko}), NGC 7129 FIRS2 is the most distant of the objects in the WISH intermediate mass sub-program. 
FIRS2 contains a 5 M$_\odot$ protostar and has a luminosity of 500 L$_\odot$ and a hot core (Fuente et al \cite{fuente01},\  \cite{fuente05a}).   
As a very young YSO it was expected that the structures dominant in this object would be those of the earliest, infalling, stage of the formation
process.
There is no evidence for a large, well-developed disk as one would expect to find in an older YSO. On the other hand, a well-developed quadrupolar outflow (due to the superposition of two bipolar outflows), has been found very close to the protostar (Fuente et al, \cite{fuente01}), indicating that some evolution past the collapsing spherical envelope stage has occurred.

A recent analysis of a number of IM YSOs included NGC7129 FIRS2 (Crimier et al \cite{crimier}). 
This analysis models the objects as spherical envelopes, includes all available far infrared ({\it Spitzer}) and submillimeter (JCMT and IRAM) brightness measurements.
The best-fit model found by Crimier et al (\cite{crimier}) has an envelope mass of 50 M$_\odot$, an optical depth at 100 $\mu$m of 2.3, an inner radius of 100 AU and an outer radius of 18,600 AU. 
The temperature at the inner envelope radius is 300 K and this falls to 100 K at a radius of 373 AU where the H$_2$ density is $4.4 \times 10^7$ cm$^{-3}$. 
The density varies as a power-law with index 1.4.

We use this model as a starting point for understanding the observations described here.
We present a first look at NGC7129 FIRS2, the youngest known IM YSO, with the PACS instrument 
(Poglitsch et al \cite{poglitsch}) 
on the {\it Herschel Space Observatory}.
The high excitation lines accessible with the instrument should provide the best probe available of the warmest, innermost regions.

\section{Observations}

The PACS instrument was used to obtain data for this project during the {\it Herschel} Science Demonstration Phase on October 26, 2009.  
The pointed line-scan mode was used to obtain data in two scans (Obs ID 1342186321 and 1342186322).  
Data were reduced using HIPE v2.4 and then analyzed using IDL. 
The relative and absolute spectral response functions are based on ground calibrations and corrected for in-flight observations by dividing fluxes below and above 100 $\mu$m by 1.3 and 1.1, respectively \footnote{From internal PACS-ICC results.}
The continuum levels were found to vary from band to band but {only a simple polynomial function was required to remove this baseline emission}.
The noise levels also varied from band to band, but a typical value was 0.3 Jy with 0.01 $\mu$m channel widths in the bands between 100 $\mu$m and 180 $\mu$m independent of scan or receiver, which corresponds well to the noise levels predicted in HSPOT, the {\it Herschel} Science Planning and Observing Tool.
The spectral resolution is 3,000 at 63.2 $\mu$m but varies from 1000 to 4000 over the wavelength range used here..
The PACS spectra are imaged in 5$\times$5 {pixels}, with each {pixel} subtending 9.4\arcsec by 9.4\arcsec, with no gaps between the pixels. For wavelengths
shortward of 110 $\mu$m this (9.4\arcsec) is the effective resolution but the psf is significantly larger than the pixels at longer wavelengths such that at 200 $\mu$m only 40\% of the light of a well-centred point source falls on the central pixel. 
{The central pixel was positioned at the coordinates of the central interferometer continuum peak at $21^h43^m1.7^s\ 66^\deg 03\arcmin 23\arcsec$ (J2000).}
While there is still an uncertainty in the flux calibration of approximately 50\%, this does not have a significant
impact on the first analysis of this object as presented here.

\begin{figure}
\centering
\includegraphics[width=8cm]{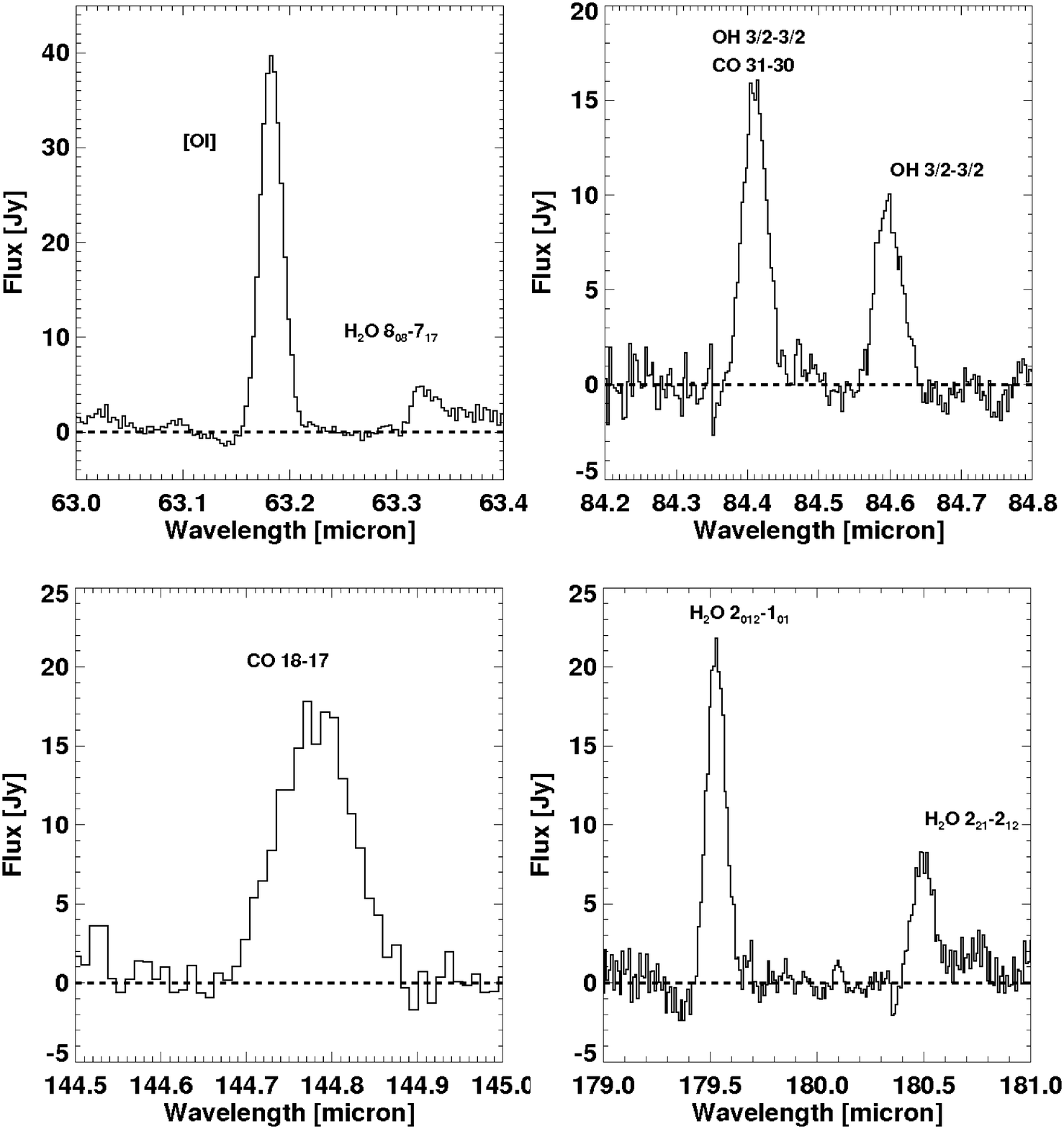}
\caption{
A sample of the spectra detected in these observations with PACS in NGC7129 FIRS 2.
}
\end{figure}

Table 1 lists 27 lines that were detected, including two pairs of blended lines.
{The fluxes given are measured only in the central pixel.}
The list of detections include 9 water lines (plus one blended with a CO line), 9 unblended CO lines, 3 OH lines (plus one blended with a CO line), and 2 [OI] lines. 
The rest wavelengths, upper level energies as well as the integrated line fluxes measured in the central PACS pixel are provided in Table 1.
A number of spectra from these observations are shown in Figure 1.  
Note that the line width is essentially all due to the instrument spectral
resolution.
Five of these water lines were covered with ISO LWS but only upper limits were detected (Crimier et al \cite{crimier}).   
In all five cases the LWS limits are a factor of ten or more greater than the detections listed here, with typical limits of $10^{-15}$ W m$^{-2}$. 

In addition to the lines listed in Table 1 there is also a detection of the [CII] line at 158 $\mu$m and numerous faint lines that can not be measured well until the calibration of PACS improves. 
[C II] 158 $\mu$m emission is detected across the entire PACS field-of-view.
{The other emission lines in Table 1 come from a compact region, consistent with being smaller than the resolution of the instrument.}
This [CII] emission is likely not directly related to the YSO but instead arises from the much larger NGC 7129 nebula itself and will not be discussed further in this paper.

\begin{table}
\caption{NGC 7129 fluxes measured from PACS spectra}

\centering
\begin{tabular}{c c c c c}
\hline
\hline
\\
Species& Transition         & $\lambda$ & E$_{\rm u}$ & Int. Flux  \\
        &                    &           &            & $10^{-18}$  \\
        &                    & $\mu$m    & K          & W m$^{-2}$  \\
\hline
\\
H$_2$O & o 2$_{12}$-1$_{01}$ & 179.5 & 114.1  & 241  \\
       & o 2$_{21}$-2$_{12}$ & 180.5 & 194.1  & 120  \\
       & o 2$_{21}$-1$_{10}$ & 108.0 & 194.1  & 295  \\
       & o 3$_{03}$-2$_{12}$ & 174.6 & 196.8  & 227  \\
       & p 3$_{13}$-2$_{02}$ & 138.5 & 204.7  & 194  \\
       & p 4$_{04}$-3$_{13}$ & 125.4 & 319.5  & 113  \\
       & o 4$_{14}$-3$_{03}$ & 113.5 & 323.5 & 596\tablefootmark{a} \\
       & p 3$_{22}$-2$_{11}$ & 90.0 & 296.8  & 113  \\
       & o 4$_{23}$-3$_{12}$ & 78.7  & 432.2  & 308\tablefootmark{b}  \\
       & o 8$_{18}$-7$_{07}$ & 63.3 & 1070.7  &  66  \\
OH     & 1/2-3/2             & 79.1  & 181    & 209  \\
       & 1/2-3/2             & 79.2  & 181    & 182  \\
       & 3/2,7/2-3/2,5/2     & 84.4  & 291    & 290\tablefootmark{c}  \\
       & 3/2,7/2-3/2,5/2     & 84.6  & 291    & 181  \\
OI     & $^3P_1$ - $^3P_2$   & 63.2  & 227.7  & 802 \\
       & $^3P_0$ - $^3P_2$   & 145.5 & 326.6  &  50\tablefootmark{d}  \\
CO     & 14-13               & 186.0   & 580.5  & 368  \\
       & 16-15               & 162.8   & 751.7  & 336  \\
       & 18-17               & 144.8 & 945.0  & 308  \\
       & 22-21               & 118.6 & 1397.4 & 273  \\
       & 24-23               & 108.8 & 1656.5 & 173  \\
       & 29-28               & 90.2 & 2400.0 & 153  \\
       & 30-29               & 87.2 & 2564.8 & 246  \\
       & 32-31               & 81.8  & 2911.2 & 144  \\
       & 33-32               & 79.4  & 3092.5 & 112  \\
\hline
\end{tabular}
\tablefoottext{a}{blended with CO (23-22)}
\tablefoottext{b}{Edge effect may have increased the flux}
\tablefoottext{c}{blended with CO (31-30)}
\tablefoottext{d}{very uncertain}
\end{table}


\section{Discussion}

The CO lines detected in these observations are all lines characteristic of very hot regions. 
CO $J=$ 33-32 for instance, arises from a level at an energy equivalent to 3093 K, {and is not likely to be strong unless the excitation temperature is in excess of 1000 K.}
The typical excitation temperatures for the other species detected here are not as high and so, for this paper, we focus on the CO lines and only briefly discuss the other observed lines. 
We first examine our original idea for the dominant structure in NGC7129 FIRS2: that this YSO is very young {and therefore is dominated by a spherical envelope.}
Is the gas, as traced by the most readily observed molecule, CO, compatible with the hypothesis of an infalling spherical envelope that can be fitted to the dust through SED modelling?
We show that this idea is untenable in the face of the data obtained and described here.
We are led to examine non-spherical geometries to explain these observations. There we examine details of the physical and radiative structure around intermediate mass protostars that could prove valuable in understanding star formation.

\subsection{CO from a spherical protostellar envelope}

The envelope model used in the analysis by Crimier et al (\cite{crimier}) provides a reasonable starting point for comparison to these new observations. 
{The envelope in their model extends to 18,600 AU or 14\arcsec at the distance of this object, close to the beam-size of these observations.}
Crimier et al (\cite{crimier}) do not make predictions for the CO lines and so we have reproduced their model envelopes and used the RATRAN radiative transfer code (Hogerheijde \& van der Tak \cite{hogerheijde}) to estimate the strength of the CO line emission.
The model results are plotted in Figure 2 as a red line showing the integrated line strength versus the upper level rotation $J_{\rm u}$.  
The CO line emission observed by PACS is shown on the same plot, indicated by data points with error bars showing the 50\% uncertainties in the measured fluxes. 

{Also shown in Figure 2 are three data points from low-J CO, integrated over a 20 arcsec wide aperture. Two CO $J$=2-1 data points are given (from Fuente, private communication), the lower one for the velocity range of -11 to -9 km/s that may be appropriate for the envelope, and the upper one for the entire range over which emission is detected (-20 to 0 km/s) appropriate to the outflow. 
Although we present the -11 to -9 km/s ambient cloud velocity value it should be noted that no peak is seen in the map in this velocity range.
The CO $J$=3-2 data point is integrated over velocities from -14 to -6.5 km/s (Miskolczi et al, \cite{miskolczi01} and most of the emission is from the outflow - indeed it is not clear that any envelope emission is present in the maps shown. A measurement of CO J=1-0 was also made from published interferometry images but the flux was too low to appear on this plot.  Single dish observations of this line show no source at this position, as with the CO J=2-1. It is not surprising that this low-J CO does not fit the spherical envelope as it likely arises from the outflow.} 

The structure in the model fit around $J_{\rm u}=10$ is due to two effects: the lines are optically thick below that point; and also we use a constant beam size of 20$\arcsec$ for $J_{\rm u}< 14$ and use the actual {\it Herschel} beam size at higher $J$.  

The predicted CO emission from the envelope for $J_{\rm u} >13$ is many orders of magnitude below the observed emission. 
This is not surprising as the volume within the protostellar envelope that exceeds 100 K is very small, with a radius of less than 400 AU. 
This radius corresponds to approximately 0.3$\arcsec$ at the distance of NGC7129. 
With a typical {\it Herschel} beamsize of 10$\arcsec$ for the CO lines, this hot region fills a small fraction (less than $10^{-3}$) of the beam area. 
We explored other possible spherical envelope models using similar constraints to the Crimier et al (\cite{crimier}) model. 
None of these alternate models have successfully produced a significant amount of CO emission in $J_{\rm u}> 15$ lines.   

\begin{figure}
\centering
\includegraphics[angle=-90,width=8cm]{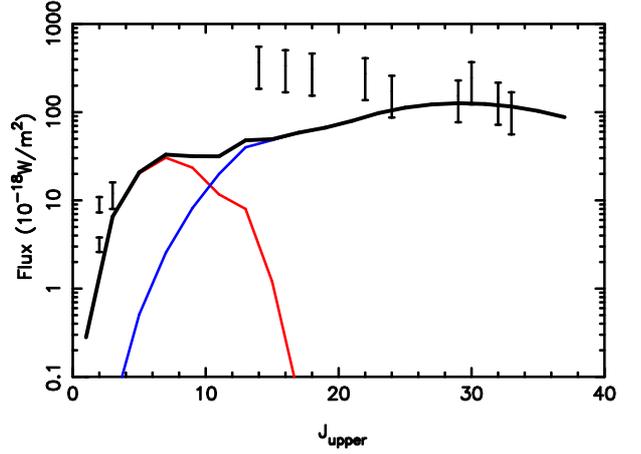}
\caption{Observed integrated CO line strengths (data points) and CO emission calculated from the protostellar envelope  model presented by Crimier et al (red line), from a CO slab model (blue line) as described in text, and the total of the flux predicted from these two models (black line). The three points at the left (two 2-1 CO and one 3-2 CO) are taken from ground-based observations as described in the text.}
\end{figure}

\subsection{CO from other components within the YSO}

A quick analysis to gain some insight into the origin of the high-$J$ CO lines is to make a Boltzmann plot (shown as Figure 3 in on-line materials). 
A single temperature LTE component (best fit at 521K) did not fit well but two components, at temperatures of 338 K and 799 K, gave a satisfactory fit.  
However temperatures up to 1300 K were found to be consistent within the uncertainties in the data.

One possible source of the high-$J$ CO lines is a hot core. However, the size of such a hot core at the distance of N7129 is small.
This hot core is equivalent to the hot inner part of the envelope described above: the size of any such core is very small (0.72" $\times$ 0.52") and has an upper temperature limit of 890 K (Fuente et al \cite{fuente05a}), and thus the CO emission will be much too small to account for this emission.

A cavity within the envelope, presumably carved by the observed outflow, heated by UV photons could present a larger hot (T $\sim$ a few 100 K) area within the beam of the PACS observations (Spaans et al. 1995 \cite{spaans95}, van Kempen et al. \cite{vankempen09}) 
{Burton et al \cite{burton90} show that strong UV emission in clumpy photodissociation models can produce significant high-J CO emission but in virtually all of their models the mid-J CO lines are several orders of magnitude stronger than the high-J CO lines, unlike what we see in this object. 
However one of their models, the low UV field, high density, ``high temperature solution'' (there are two possible solutions) can produce high-J lines nearly as strong as the mid-J lines.  The Burton et al models also predict very much stronger emission in lines near J=10-9, providing a test of whether or not this model applies to this object.}

While UV heated cavity walls can likely produce the mid-$J$ CO line emission that we see (see the similar case for HH46, van Kempen et al \cite{vankempen10}), the highest-$J$ CO lines probably require still hotter (T $>$ 1,000 K) gas and we suggest that this gas is heated by non-dissociative shocks.
Here we use the more general RADEX code (van der Tak et al \cite{vandertak07}) to model CO emitting slabs at the temperatures and densities which other studies have shown are appropriate for these mechanisms.
A detailed analysis to explain the CO that considers many other issues will be carried out in a future paper. 

Figure 2 shows the results of one such model (blue line).  
We present this as an illustration of how this CO emission can arise and stress this is not a fit to the data.
Plotted in Figure 2 are the results from a RADEX model of the CO emission from a plane parallel slab with a kinetic temperature of 1100 K, an H$_2$ density of $1.0\times 10^8$ cm$^{-3}$, and a CO column density of $1.0\times 10^{14}$ cm$^{-2}$.  This temperature is similar to the one derived towards low-mass protostars by Neufeld et al \cite{neufeld09}.
The line width (10 km/s) does not affect the result at these low optical depths (typically $\tau = 2\times 10^{-5}$ in these CO lines).
 Note that, as with the envelope model, the emission is integrated over a beamsize of 15$\arcsec$ for  $J_{\rm u} <14$; a beam area appropriate to the diffraction limited {\it Herschel} beam, is used up to $J_{\rm u}=20$. However for the higher $J$ lines (that do not appear in the envelope model) the PACS beam of 9.4$\arcsec$ is used.
The mass contained within the beam, assuming a CO abundance of $10^{-4}$, is negligible (of order $10^{-5}$ M$_\odot$).

\onlfig{3}{
\begin{figure}
\includegraphics[angle=-90,width=16cm]{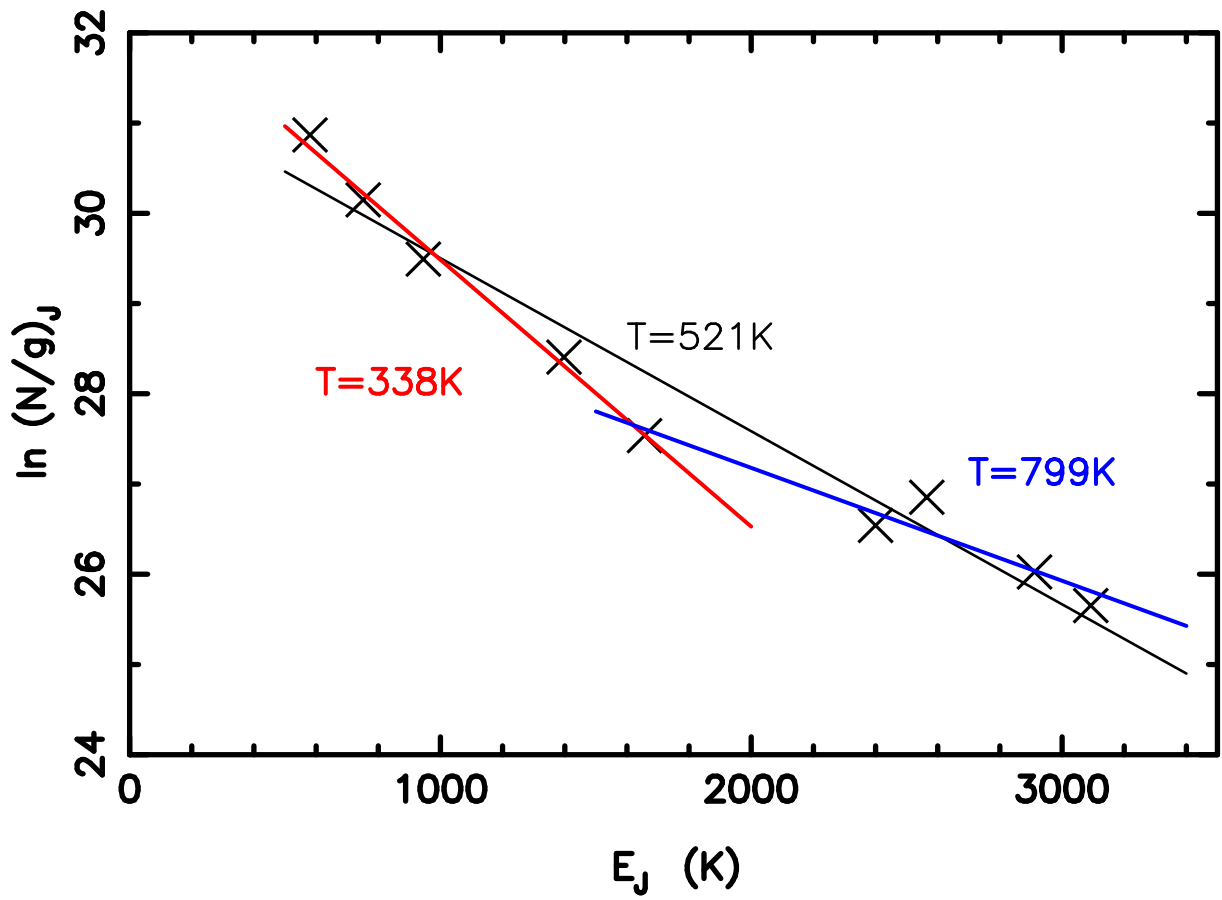}
\caption{A Boltzmann plot of the CO emission lines observed with PACS. 
This plot shows the column density versus the excitation energy of the 
upper level.
The best fit to a single component gives a temperature of 521 K but the 
fit is not good. Two components with temperatures of 338 K and 799 K 
produces a better fit, as shown.}
\end{figure}
}

This slab model provides a reasonable and simple first iteration fit to the observed CO fluxes.
The highest-$J$ CO lines provide an excellent match between the observed and predicted fluxes. 
The observed fluxes of the mid-$J$ lines are much greater than the modeled fluxes, by a factor of eight for the first of these lines ($J$=14-13).  
We have not found a single CO component that can match the observed flux from these three lines without simultaneously producing significantly too much flux in highest lines.  
It appears likely that at least two separate components are required within the current plane-parallel slab geometry in order to match all of the observed lines.
With further calibration of the PACS instrument, the uncertainties in the observed line strengths are expected to diminish. 
As well, this source will be observed with HIFI as part of the WISH project. 
If the mid-$J$ CO lines observed are from shocks then HIFI observations should resolve the lines and show them to be broad.
A more detailed consideration of the structure within the protostellar envelope will be best undertaken once these additional data and calibrations are available.

We defer a detailed discussion of the other molecules detected to future papers.
The H$_2$O data listed in this paper will be combined with recently obtained HIFI data in one of these future papers.


\section{Conclusions}

The principal result of these new observations by  PACS is that the unexpected strength of the observed high excitation (high-$J$) CO lines requires high temperatures as well as relatively high densities to be present in the immediate vicinity of NGC 7129 FIRS2.
We have found a single component plane parallel hot slab can produce a reasonable fit to the highest-$J$ CO observations but a more complex solution is necessary to fit all the lines simultaneously, especially the mid-$J$ lines.
This hot component may represent a UV photon heated inner wall of the outflow cavity or may be related to shocks associated with the outflow itself. 
In the future it may be possible to model all the CO emission using a self-consistent 2-D geometric shock and UV heating model as has been used in the interpretation of the observations of the low mass protostar HH46 (van Kempen et al \cite{vankempen10}), where similar highly excited CO emission is seen. 
HIFI observations will provide  the higher spectral resolution needed to trace the kinematics of these structures.

\begin{acknowledgements}
We thank Bruno Merin, Jeroen Bouwman, and Bart VandenBussche of the PACS ICC for all of their help with the data reduction.
JC and AF thank theo Spanish MCINN for funding support under program
CONSOLIDER INGENIO 2010 ref: CSD2009-00038, and JC, under programs
AYA2006-14786 and  AYA2009-07304.
A portion of this research was performed at the Jet Propulsion Laboratory, California Institute of Technology, under contract with the National Aeronautics and Space Administration.
This program is made possible thanks to
the HIFI guaranteed time program and the PACS instrument builders.
HIFI has been designed and built by a consortium of institutes and university departments from across Europe, Canada and the United States under the leadership of SRON Netherlands Institute for Space Research Groningen, The Netherlands and with major contributions from Germany, France, and the US. Consortium members are: Canada: CSA, U.Waterloo; France: CESR, LAB, LERMA, IRAM; Germany: KOSMA, MPIfR, MPS; Ireland: NUI Maynooth; Italy: ASI, IFSI-INAF, Osservatorio Astrofisico di Arcetri-INAF; Netherlands: SRON, TUD; Poland: CAMK, CBK; Spain: Observatorio Astronomico Nacional (IGN), Centro de Astrobiologia (CSIC-INT); Sweden: Chalmers University of Technology - MC2, RSS \& GARD, Onsala Space Observatory, Swedish National Space Board, Stockholm University - Stockholm Obseratory; Switzerland: ETH Zurich, FHNW; USA: Caltech, JPL, NHSC.
\end{acknowledgements}

\end{document}